\documentclass[structabstract]{aa}  
\usepackage{natbib}	
\usepackage{graphicx}
\usepackage{txfonts,epsfig,graphicx,url,twoopt}
\usepackage[breaklinks=true]{hyperref} 
\hypersetup{colorlinks=true,citecolor=blue}
\bibpunct{(}{)}{;}{a}{}{,}     
\usepackage[usenames,dvipsnames]{color}

\begin{document}
\title{Explaining millimeter-sized particles in brown dwarf disks}
\author{P.~Pinilla\inst{1,2}, T.~Birnstiel\inst{3,4}, M.~Benisty\inst{5}, L.~Ricci\inst{6}, A.~Natta\inst{7,8}, C.~P.~Dullemond\inst{1}, C.~Dominik\inst{9,10} and L.~Testi\inst{7,11}}

\institute{Universit\"at Heidelberg, Zentrum f\"ur Astronomie, Institut f\"ur Theoretische Astrophysik, Albert-Ueberle-Str. 2, 69120 Heidelberg, Germany\\
\email{pinilla@uni-heidelberg.de}
\and
Member of IMPRS for Astronomy \& Cosmic Physics at the University of Heidelberg
\and
Harvard-Smithsonian Center for Astrophysics, 60 Garden Street, Cambridge, MA 02138, USA
\and
Excellence Cluster Universe, Boltzmannstr. 2, D-85748 Garching, Germany
\and
Laboratoire d'Astrophysique, Observatoire de Grenoble, CNRS/UJF UMR 5571, 414 rue de la Piscine, BP 53, 38041 Grenoble Cedex 9, France
\and
California Institute of Technology, MC 249-17, Pasadena, CA, 91125, USA
\and
INAF - Osservatorio Astrofisico di Arcetri, Largo Fermi 5, 50125, Firenze, Italy
\and
School of Cosmic Physics, Dublin Institute for Advanced Studies, 31 Fitzwilliam Place, Dublin 2, Ireland
\and
Astronomical Institute ÒAnton PannekoekÓ, University of Amsterdam, PO Box 94249, 1090 GE Amsterdam, The Netherlands
\and
Department of Astrophysics/IMAPP, Radboud University Nijmegen, PO Box 9010, 6500 GL Nijmegen, The Netherlands
\and
European Southern Observatory, Karl Schwarzschild Str. 2, D-85748 Garching bei M\"unchen, Germany}

\date{Received 7 December 2012; Accepted 22 April 2013}

%%%%%%%%%%%%%
% ABSTRACT
%%%%%%%%%%%%%
 
  \abstract
   % context heading (optional)
{Planets have been detected around a variety of stars, including low-mass objects, such as brown dwarfs. However, such extreme cases are challenging for planet formation models. Recent sub-millimeter observations of disks around brown dwarf measured low spectral indices of the continuum emission that suggest that dust grains grow to mm-sizes even in these very low mass environments.}
   % aims heading (mandatory)
   {To understand the first steps of planet formation in scaled-down versions of T-Tauri disks, we investigate the physical conditions that can theoretically explain the growth from interstellar dust to millimeter-sized grains in  disks around brown dwarf.}
  % methods heading (mandatory)
   {We modeled the evolution of dust particles under conditions  of low-mass disks around brown dwarfs. We used coagulation, fragmentation and disk-structure models to simulate the evolution of dust, with zero and non-zero radial drift. For the non-zero radial drift, we considered strong inhomogeneities in the gas surface density profile that mimic long-lived pressure bumps in the disk. We studied different scenarios that could lead to  an agreement between theoretical models and the spectral slope found by millimeter observations.}
  % results heading (mandatory)
   {We find that fragmentation is less likely and rapid inward drift is  more significant  for particles in brown dwarf disks than in T-Tauri disks. We present different scenarios that can nevertheless explain millimeter-sized grains. As an example, a model that combines  the following parameters  can fit the millimeter fluxes measured for brown dwarf disks:  strong pressure inhomogeneities of $\sim$~40\% of amplitude, a small radial extent $\sim$~15~AU, a moderate turbulence strength $\alpha_{\mathrm{turb}}~=~10^{-3}$, and average fragmentation velocities for ices $v_f~=~10~m~s^{-1}$.}
{}
  

\authorrunning{Pinilla P. et al.}

\maketitle

%%%%%%%%%%
\section{Introduction}     \label{sec:intro}
%%%%%%%%%%
Since the first confirmed discovery of brown dwarf (BD) \emph{Teide~1}  \citep{rebolo1995} and \emph{Gliese~229B} \citep{nakajima1995} in 1995, several hundred  BDs  have been identified and many efforts have  focused on understanding these objects which are considered to be an intermediate step between planets and stars.  
Observations of BD show near-infrared excess emission \citep[e.g.][]{muench2001, liu2003} that reveals material around young BDs. Moreover, typical fluxes measured with millimeter observations \citep[e.g][]{klein2003, scholz2006, joergens2012} are in most of the cases lower than   few mJy at 1~mm, implying that these circumstellar disks have masses of few $M_{\mathrm{Jup}}$ or even lower.  Determining whether these low-mass disks can be the scene of the formation of planetesimals or even planets is still a subject of debate.

Observationally, some aspects of disks around BD are different from those around T-Tauri and Herbig Ae/Be stars. They have a  lower accretion rate \citep[$\thicksim~10^{-12}~M_{\odot}~yr^{-1}$,][]{herczeg2009}, a flat tendency for the disk geometry inferred from \emph{Spitzer} observations and spectral energy distribution (SED) modeling \citep[e.g.][]{apai2005, allers2006, guieu2007, scholz2007, morrow2008, pascucci2009, szucs2010},  and a longer lifetime \citep{carpenter2006, riaz2012, harvey2012}.

To study dust growth in protoplanetary disks, different  mechanisms should be taken into account, as for instance turbulent mixing, settling, aerodynamical drag with the gas, collision rates and fragmentation.  Essentially, when the dust particles are small, they are well coupled to the gas,  move along with it, and  grow as a consequence of surface forces. However, when the particles increase in size, they start to  decouple from the gas and the relative velocities between particles increase, leading to fragmentation collisions \citep{weidenschilling1977, brauer2008}. In addition, before any meter-sized object can be formed at  Earth-Sun distances, dust drifts toward the central star because of the sub-Keplerian velocity of the gas. Millimeter (mm) grains experience the same rapid inward migration in the outer regions of the disks, even though mm-size particles have been observed in those  regions of protoplanetary disks \citep[e.g.][]{wilner2000, natta2004, rodmann2006, ricci2010a, ricci2010b, ricci2011, ubach2012}. Local pressure maxima in disks have been proposed as a solution of this rapid inward drift \citep{klahr1997, johansen2009, pinilla2012a}.

\cite{birnstiel2010b} showed that under typical T-Tauri parameters, neglecting radial drift and considering different parameters for the disk such as  turbulence,  particles could reach millimeter sizes, meaning  that the spectral index $\alpha_{1-3\mathrm{mm}}$ can have low values.  \cite{birnstiel2010b}  models predict that for a disk mass lower than 5~$M_{\mathrm{Jup}}$, such as the mass of a BD disk, $\alpha_{1-3\mathrm{mm}}$ would be close to 3. However, recent millimeter observations confirmed low values  $\alpha_{1-3\mathrm{mm}}$ for  two BD disks \citep{ricci2012, ricci2013}, with $\alpha_{1-3\mathrm{mm}}~\approx$~2.3. In addition, as we show below, the radial drift and fragmentation  barriers are  different for particles in disks around BD than for those around  more massive and luminous stars.  Radial drift indeed has a stronger influence on particles in BD disks, and as a consequence, any mechanism in a disk  that may allow a reduced inward migration of grains has to be more extreme in BD disks to lead to  an effective trapping of particles. Explaining how the first pebbles  are formed from interstellar dust in BD disks is therefore a very intriguing topic.

The purpose of this paper is to investigate whether the dust growth models that were successful for T-Tauri disks are, when applied to BD disks, consistent with a short set of observations at millimeter wavelengths (two measurements from recent ALMA  and CARMA observations). As mentioned above, infrared studies have on the other hand focused on larger surveys, showing interesting trends such as a flatter BD disk geometry. However, we do not aim to reproduce these trends, because infrared observations yield no strong constraints on the dust evolution models (other than those studied in e.g. \cite{szucs2010} or \cite{mulders2012}).

For this work, dust coagulation/fragmentation models were considered in two main cases: first, the extreme case of setting the radial drift to zero, and second, taking radial drift into account with strong inhomogeneities in the gas surface density  that mimic long-lived pressure bumps. In both cases, different disk parameters were considered, to analyze which scenarios are the best incubators of the first pebbles found in BD disks with millimeter observations. In Sect.~\ref{sec2}, we describe the drift and fragmentation barriers for the specific case of BD disks and the physical parameters of the dust coagulation/fragmentation model. Numerical results, observational perspectives, and comparisons with current mm-observations are presented in Sect~\ref{sec:results}. A summary of the results with the corresponding discussion of this work is  recapitulated in Sect. \ref{sec:discussion}. Finally, the main conclusion of this paper is given in Sect.~\ref{sec:conclusion_BD}.

%%%%%%%%%%
\section{Dust evolution model, drift, and fragmentation barriers in BD disks}     \label{sec2}
%%%%%%%%%%

The interaction between the gas and the dust is fundamental for understanding how the particles evolve within the protoplanetary disk. When particles are well coupled to the gas, the dust relative velocities are mainly generated by Brownian motion and settling to the midplane. Considering these two sources for the velocities, particles stick by van der Waals forces and collisional growth is very efficient, producing mm-size grains in the outer regions ($\gtrsim~$50~AU)  on timescales of $\sim~10^{5}$ years in typical T-Tauri disks \citep{birnstiel2010a}. Unfortunately, when particles grow, the relative velocities substantially increase through turbulent motion and radial drift; the collision energies are therefore high enough to cause fragmentation, as was experimentally shown by \cite{blum2008}. From the theoretical point of view, stellar and disk properties strongly influence these first steps of planet formation, meaning that effects such as the usual inward migration of dust grains may vary for BD disks. In this section, we first explain the main characteristics of the numerical model for the dust evolution, followed by an explanation of  the drift  and fragmentation barriers in BD disks, and finally we describe the set-up for the numerical simulations that are considered in this work.

\subsection{Dust evolution model} \label{sec2_1}
For the dust evolution, we used the coagulation/fragmentation model explained in \cite{birnstiel2010a}. The dust evolution is described by the  advection-diffusion differential equation of the dust surface density $\Sigma_d$, which for a single dust size can be written in cylindrical coordinates as

\begin{equation}
	\frac{\partial \Sigma_d}{\partial t} + \frac{1}{r}\frac{\partial}{\partial r}\left( r \Sigma_d u_{\mathrm{r,d}}\right)-\frac{1}{r}\frac{\partial}{\partial r} \left(r \Sigma_g D_d \frac{\partial }{\partial r}\left[\frac{\Sigma_d}{\Sigma_g}\right]\right)=0,
  \label{eq:dustevo}
\end{equation}

\noindent where $D_d$ is the dust diffusivity and $\Sigma_g$ is the gas surface density. Because the timescales for gas viscous evolution are longer than the dust growth timescales, we considered that for the dust evolution models, the gas surface density remains constant with time. This equation was solved for each size using the flux-conserving donor-cell scheme \citep[see][Appendix A]{birnstiel2010a}. The radial velocity of the dust $u_{\mathrm{r,d}}$ has two contributions: the first one $u_{\mathrm{drag}}$ comes from the drag with the gas that depends on the radial gas velocity $u_{\mathrm{r,g}}$ and on the size of the particles; and the second one arises from radial drift $u_{\mathrm{drift}}$, which is proportional to the radial pressure gradient $\partial_r P$, such that

\begin{equation}
	u_{\mathrm{r,d}}=\frac{u_{\mathrm{r,g}}}{1+\textrm{St}^2}+\frac{1}{\textrm{St}^{-1}+\textrm{St}} \frac{\partial_r P}{\rho_g \Omega},
 \label{eq:dustvel} 
\end{equation}

\noindent where the coupling constant St -the Stokes number- is defined as the ratio between the largest eddy turn-over time ($1/\Omega$, with $\Omega~=~\sqrt{G~M_{\star}~r^{-3}}$) and the stopping time of the particle within the gas. In the Epstein regime,  where the ratio between the mean free path of the gas molecules $\lambda_{\mathrm{mfp}}$ and the sizes of the particles $a$ is $\lambda_{\mathrm{mfp}}/a~\geq~4/9$, St is defined at the disk midplane as

\begin{equation}
	\textrm{St}=\frac{a\rho_s}{\Sigma_g}\frac{\pi}{2},
  	\label{eq:stokes}
\end{equation}

\noindent where $\rho_s$ is the volume density of a dust grain, usually of about $\sim$~1~g~cm$^3$. 

In addition, the turbulent gas viscosity is considered as $\nu~=~\alpha_{\mathrm{turb}}~c_s^2~\Omega^{-1}$ \citep{shakura1973}, where $c_s$ is the isothermal sound speed, which is given by

\begin{equation}
	c_s^2= \frac{k_B T(r)}{\mu m_p},
  	\label{eq:sound} 
\end{equation}

%%%%%%%%%%%%
%FIGURE: DUST DISTRIBUTION WITH DRIFT R=30 vf=10
%%%%%%%%%%%%
\begin{figure*}
   \centering
   	\includegraphics[width=18.0cm]{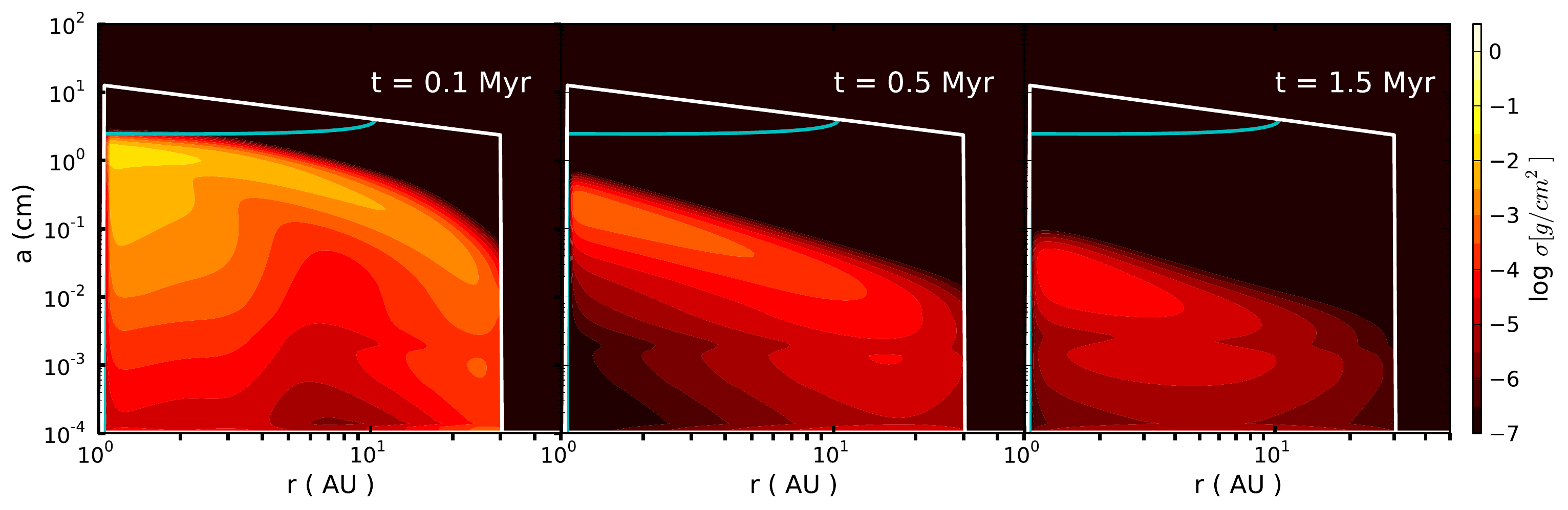}
   \caption{Vertically integrated dust density distribution (Eq.~\ref{eq:vertically}) after different times of evolution and including radial drift. Case of the BD disk $\rho$-Oph102 parameters ($M_{BD}~=~0.05~M_\odot$, $L_{BD}~=~0.03~L_\odot$ and  $T_{BD}~=~2880~K$) and $R_{\mathrm{out}}~=~30$~AU, $\Sigma~=~\Sigma_0~r^{-0.5}$, $M_{\mathrm{disk}}~=~2~M_{\mathrm{Jup}}$, $v_f~=~10~m~s^{-1}$ and $\alpha_{\mathrm{turb}}~=~10^{-3}$. The solid white line represents the particle size corresponding to St~=~1 (Eq.~\ref{eq:stokes}) and reflects the shape of the gas density. The cyan line corresponds to the largest size that particles can reach given a fragmentation velocity $v_f$.}
   \label{with_drift}
\end{figure*}

\noindent where $k_B$ is the Boltzman constant, $m_p$ the proton mass and $\mu$ the mean molecular mass. Magnetorotational instability (MRI) is the most likely source of turbulence in disks \citep[e.g][]{{johansen2005}}. MRI essentially depends on the disk temperature and penetration of cosmic, X- and UV-rays to the midplane. The dust diffusivity in Eq.~\ref{eq:dustevo} can be defined in terms of St when the gas diffusivity is considered to be the turbulent gas viscosity $\nu$ \citep{youdin2007}, hence

\begin{equation}
	D_d=\frac{\nu}{1+\mathrm{St}^2}.
  	\label{eq:diffusion}
\end{equation}

The Stokes number St describes how well  the particles are coupled to the gas. When St~$\ll$~1, the first term dominates in Eq~\ref{eq:dustvel}, i.e., $u_{\mathrm{drag}}$, and as a result these particles move along with the gas, meaning that they have sub-Keplerian velocities. When the particle size increases, the second term, i.e., $u_{\mathrm{drift}}$ starts to dominate and reaches the highest  value when St~=~$1$, and as a consequence these are the particles that react strongest  to the sub-Keplerian velocity of the gas. The bodies with St~$\gg$~1 move with their own velocity, i.e., Keplerian speed.

In addition, dust particles grow, fragment, and crater depending on the relative velocities between them. For this the Smoluchowski coagulation equation \citep{smoluchowski1916} was solved  for the dust grain distribution $n(a,r,z)$ \citep[see][Eq.~35 and Eq.~36]{birnstiel2010a}, considering three different physical processes: coagulation, fragmentation, and erosion. For the relative velocities, we assumed Brownian motion, settling, turbulent motion \citep{ormel2007}, and drift velocities in the azimuthal and radial direction. The fragmentation velocities $v_f$ were estimated based on  laboratory experiments and theoretical work of collisions, which are of the order of few $\mathrm{m~s}^{-1}$  for silicates \citep{blum2008} and several $\mathrm{m~s}^{-1}$ for ices  \citep[e. g.,][]{wada2009, wada2011}. To describe the dust grain distribution, we  refer to the vertically integrated dust surface density distribution per logarithmic bin, which is given by

\begin{equation}
	\sigma (r,a)=\int_{-\infty}^{\infty} n(r,z,a)\cdot m\cdot a dz,
  \label{eq:vertically}
\end{equation}

\noindent where $m$ is the mass of a single particle of size $a$. Therefore the total dust surface density can be written as

\begin{equation}
	\Sigma_d(r)=\int_0^\infty \sigma (r,a)d\ln a.
  \label{eq:Sigmadust}
\end{equation}

\subsection{High radial drift and fragmentation in BD disks}  \label{sec2_2}

In a protoplanetary disk, large bodies feel no pressure and move with Keplerian speed, while the gas is slightly  sub-Keplerian. As a result, particles that have grown to a size where they start to decouple from the gas, i.e., St~$\approx$~1, feel a strong headwind, lose angular momentum, and move inward. Solving the radial Navier-Stokes equation, the azimuthal velocity of the gas is given by \citep{nakagawa1986}

\begin{equation}
	u_{\phi,\mathrm{g}}=u_k\left(1-\eta\right),
  	\label{eq:gas_vel} 
\end{equation}

\noindent where $u_k$ is the Keplerian speed and

\begin{equation}
	\eta=-\frac{1}{2 \rho_g(r,z) r \Omega^2}\frac{dP(r)}{dr},
  	\label{eq:eta} 
\end{equation}

\noindent with  $\rho_g(r,z)$ the gas density, such that the gas surface density is given by $\Sigma_g=\int^{\infty}_{\infty}\rho_g(r,z) dz$. For an ideal gas, the pressure $P(r)$ is defined as

\begin{equation}
	P(r)=c_s^2 \rho_g(r,z),
  	\label{eq:pressure} 
\end{equation}

\noindent and therefore

\begin{equation}
	\eta=-\frac{c_s^2}{2 u_k}\left(\frac{d \ln \rho}{d \ln r} + \frac{d \ln c_s^2}{d \ln r} \right),
  	\label{eq:etaeta} 
\end{equation}

\noindent where the term in parenthesis depends only on the exponents that characterize the power-law gas density and  temperature  radial profiles in a flared disk. For the temperature, $T(r)$  can be approximated by \citep{kenyon1987}

\begin{equation}
	T(r)=T_{\star}\left(\frac{R_{\star}}{r}\right)^{1/2} \alpha_{\mathrm{inc}}^{1/4} \propto \frac{L_{\star}^{1/4}}{\sqrt{r}},
  	\label{eq:temperature} 
\end{equation}

\noindent where $r/R_\star \gtrsim 10^2$. $\alpha_{\mathrm{inc}}$ is the angle between the incident radiation and the local disk surface. For the last proportionality of Eq.~\ref{eq:temperature} the weak dependence of  $\alpha_{\mathrm{inc}}$  with $R_\star$ is neglected. Considering Eq.~\ref{eq:temperature} for the disk temperature and assuming a simple power-law function for the gas density profile, the difference between the orbital gas velocity and the Keplerian speed scales as (independently of the radial location $r$)

\begin{equation}
	u_{\phi,\mathrm{g}} -u_k=\frac{c_s^2}{2 u_k}\left(\frac{d \ln \rho}{d \ln r} + \frac{d \ln c_s^2}{d \ln r} \right)\propto \frac{T(r)}{\sqrt{M_{\star}r^{-1}}}\propto\frac{L_{\star}^{1/4}}{\sqrt{M_{\star}}}.
  	\label{eq:eta2} 
\end{equation}

This difference determines how much angular momentum the particles lose, and as a consequence, how fast they drift to the central star.  Taking Eq.~\ref{eq:dustvel} for the dust radial velocity, and considering grains with St~$\gtrsim~1$, the drag term (first term of Eq.~\ref{eq:dustvel}) can be neglected and $u_{\mathrm{r,d}}$ can be written in terms of the difference $u_{\phi,\mathrm{g}} -u_k$ as

\begin{equation}
	u_{\mathrm{r,d}}=2\frac{u_{\phi,\mathrm{g}} -u_k}{\textrm{St}^{-1}+\textrm{St}}.
	\label{eq:uradial2} 
\end{equation}

For comparison, taking the parameters of the BD know as $\rho-$Oph~102 \citep{ricci2012} and comparing them with a young solar-type star, i.e., $M_{\rho\mathrm{-Oph}102}~=~0.05~\times~M_{\mathrm{sun}}$ and $L_{\rho\mathrm{-Oph}102}~=~0.03~\times~L_{\mathrm{sun}}$, this leads to

\begin{equation}
	\left(u_{\phi,\mathrm{g}} -u_k\right)_{\rho\mathrm{-Oph}102}\approx2\left(u_{\phi,\mathrm{g}} -u_k\right)_{\mathrm{sun}}.
  	\label{eq:comparison} 
\end{equation}

The fact that the radial drift is higher for BD disks than for Sun-like disks also implies that destructive collisions due to radial drift are more likely because it contributes to increase the relative velocities between the particles. Therefore rapid inward drift and the potential  fragmentation due to these high drift velocities are major problems for particles in BD disks.

In addition, the highest turbulent relative velocity of particles with a given Stokes number is given by \citep{ormel2007}

\begin{equation}
	\Delta u_{\mathrm{max}}^2\simeq \frac{3~\alpha_{\mathrm{turb}}}{\mathrm{St}+\mathrm{St}^{-1}}c_s^2.
  \label{eq:u_turb}
\end{equation}

When the radial drift is set to zero, the largest size of particles $a_{\mathrm{max}}$ is  calculated when the turbulent relative velocities (Eq.~\ref{eq:u_turb}) are equal  to the fragmentation velocities $v_f$. For particles with St~$\lesssim$~1, $a_{\mathrm{max}}$ is approximately given by

\begin{equation}
	a_{\mathrm{max}}\approx\frac{2 \Sigma_g}{3 \pi \alpha_{\mathrm{turb}}\rho_s}\frac{v_f^2}{c_s^2}.
	\label{eq:amax} 
\end{equation}

This implies that  considering only turbulent velocities, $a_{\mathrm{max}}$ will depend on disk parameters such as $\alpha_{\mathrm{turb}}$, $\Sigma_g$, and $T(r)$, which strongly vary between BD and T-Tauri disks. Whether turbulence or radial drift is the cause for destructive collisions for the dust within the disk, fragmentation would occur differently for BD disks, as we discuss in the following sections. 

As a general illustration of this problem in BD disks, Fig.~\ref{with_drift} shows the vertically integrated dust density distribution (Eq.~\ref{eq:vertically}) after different times of evolution, including radial drift and using the dust evolution model described in Sect.~\ref{sec2_1}. For this case, taking the parameters $\rho$-Oph 102 ($M_{BD}~=~0.05~M_\odot$, $L_{BD}~=~0.03~L_\odot$ and  $T_{BD}~=~2880~K$), a truncated power law is taken for the gas density profile with   $\Sigma~=~\Sigma_0~r^{-0.5}$ and $r~\in~(1,30)$~AU such that the mass of the disk is $M_{\mathrm{disk}}~=~2~M_{\mathrm{Jup}}$.  The largest size that particles can reach is represented by the solid cyan line, which was computed considering the fragmentation velocity as $v_f~=~10~m~s^{-1}$ and the turbulence parameter as $\alpha_{\mathrm{turb}}~=~10^{-3}$. The solid white line corresponds to St~=~1, which reflects the shape of the gas surface density based on Eq.~\ref{eq:stokes} and particles that feel the strongest radial drift. The dust particles initially grow in the disk, allowing to have mm size grains in the outer regions ($r~\gtrsim~10$~AU) after 0.1~Myr of evolution (left panel of Fig.~\ref{with_drift}). However, when particles grow to sizes close to St~$\sim~1$, turbulence and radial drift  lead to fragmentation and inward migration of particles, and in only  0.5~Myr of evolution (middle panel of Fig.~\ref{with_drift}), the outer region is  empty of mm-grains. This scenario does not change significantly with time (right panel of Fig.~\ref{with_drift}), but  the inner region ($r~\lesssim~10$~AU) is even more depleted of mm grains, since they continue fragmenting and drifting toward the star, leading to a dust-poor disk after 1.5~Myr. As a result, explaining how micron-size dust grows to pebbles and how those are retained in the outer regions of BD disks is  very  challenging.

In addition, if perfect sticking and  micron-sized compact particles  are considered to be formed, the mean growth time of monomers with mass $m$ at a fixed distance from the star is \citep{brauer2008, okuzumi2011}

\begin{equation}
	\tau_{\mathrm{grow}}=\left(\frac{1}{m} \frac{dm}{dt}\right)^{-1},
  	\label{eq:Tgrow} 
\end{equation}

\noindent where $dm/dt=\rho_d \sigma_{\mathrm{coll}} \Delta v$, with $\rho_d$ is the dust density, $\sigma_{\mathrm{coll}}$ the collision cross-section and $\Delta v$ the collision velocity.  For a narrow-size distribution, perfect mixing and settling assumed in the Epstein regime,  a given dust-to-gas ratio, the growth timescale is proportional to

\begin{equation}
	\tau_{\mathrm{grow}}\propto \frac{\Sigma_g}{\Sigma_d}\Omega^{-1},
  	\label{eq:Tgrow2} 
\end{equation}
 
\noindent with $\Sigma_d$ as the dust surface density. The growth timescale in disks of BD and T-Tauri disks can be compared considering, for example, regions of similar temperature.  Following \cite{mulders2012}, we re-scaled the distances with the stellar luminosity as well as  the Keplerian frequency, to obtain

\begin{equation}
	\tau_{\mathrm{grow}}^{\mathrm{BD}}\propto \frac{\Sigma_g}{\Sigma_d}\Omega_{\mathrm{BD}}^{-1}=\tau_{\mathrm{grow}}^{\mathrm{TT}}\sqrt{\frac{L_{\mathrm{BD}}}{L_\star}} ,
  	\label{eq:Tgrow3} 
\end{equation}

\noindent where $\tau_{\mathrm{grow}}^{\mathrm{BD}}$ and $\tau_{\mathrm{grow}}^{\mathrm{TT}}$ correspond to the growth timescales in BD and T-Tauri disks respectively. This implies that for micron-size particles and considering for simplicity only settling motion, the mean collision time $\tau_{\mathrm{grow}}$  is generally shorter for BD than for T-Tauri disks. As an illustration, if we again  take the BD  $\rho-$Oph~102, the growth timescale due to settling is approximately one order of magnitude shorter for dust particles around this BD than for particles in a Sun-like disk. As a consequence of the result that settling and radial drift occur faster in BD disks,  the very early stages of dust growth, when particles just stick and grow through molecular interactions, is more efficient at the location of a given temperature in BD disks than in T-Tauri disks.

Based on the low spectral indices ($\alpha_{\mathrm{1-3mm}}<3$) measure with recent ALMA-Cycle~0-observations at 3.2~mm and 0.89~mm of the young BD $\rho-$Oph~102 \citep{ricci2012}  and  CARMA observations  of the BD disk 2M0444+2512 at 3~mm \citep{ricci2013} and 0.850~mm \citep{bouy2008}, we aim to explain how mm-grains can form and be retained in disks around BD. We divide this work in two parts: In the first part, we do not allow particles to drift to study whether we can create mm-grains considering only grain growth and fragmentation. In the second part,  we aim to explain how these grains are trapped in the outer regions of BD disks when radial drift is included.

\subsection{Set-up}\label{sec2_3}
%
%%%%%%%%%%%%
%TABLE 1
%%%%%%%%%%%%
\begin{table}
\caption{Model parameters}    
\label{table_parameters}     
\centering                         
\begin{tabular}{c c }       
\hline\hline                 
Parameter & Values \\    
\hline 
   $M_{BD} [M_\odot]$ & $0.05$  \\
   $L_{BD} [L_\odot]$ & $0.03$  \\  
   $T_{BD}$ [K]&$2880$\\ 
    $\rho_s$ [g/cm$^3$]& $1.2$\\
   $M_{\mathrm{disk}} [M_{\mathrm{Jup}}]$ & $2.0$  \\      
   $R_{\mathrm{in}}$[AU] &1.0\\                   
   $R_{\mathrm{out}}$[AU] &$\{15,30,60,100\}$\\
   $\alpha_{\mathrm{turb}}$ & $\{10^{-5}, 10^{-4}, 10^{-3} \}$  \\  
   $v_f$ [m/s]& $\{10, 30\}$\\
   $A$& $\{0.4, 0.6\}$\\
   $f$& 1\\
\hline                     
\end{tabular}  
\end{table}

For the simulations of this work, we considered two different scenarios.  In the first case, there is no dust radial evolution, i.e., the drag and drift terms in Eq.~\ref{eq:dustvel} are neglected. In the second case, we considered dust radial evolution with pressure bumps.  For the gas density, we considered truncated power-law functions $\Sigma_g~=~\Sigma_0~r^{-p}$, with $p~=~\{0.0, 0.5, 1.0\}$ and $\Sigma_0$ was computed such that the mass of the disk was always $M_{\mathrm{disk}}~=~2~M_{\mathrm{Jup}}$. For the outer radius of the disk, four possibilities were considered: 15, 30, 60 and 100~AU, since the exact typical disk spatial extent is unknown  \citep{luhman2007, ricci2012}. 

The turbulence parameter $\alpha_{\mathrm{turb}}$ was taken to have values between $10^{-5}-10^{-3}$. This parameter directly influences  the maximum size of particles (Eq.~\ref{eq:amax}), if turbulence is the cause of fragmentation.  \cite{mulders2012} inferred $\alpha_{\mathrm{turb}}\sim 10^{-4}$ from SED modeling, assuming   a fixed grain size distribution and gas-to-dust ratio. The fragmentation velocities were assumed to be $v_f~=~\{10, 30\}$~m~s$^{-1}$, which are the values expected for ices \citep{wada2009}. For all  simulations, the dust-to-gas ratio was initially considered to be 1\%, the initial size of the particles was taken to be 1~$\mu m$ and the maximum size that particles can reach in the simulations was fixed to be 1~km, because for these sizes gravitational effects start to play a role, which are not included in the present dust evolution model.

%%%%%%%%%%%%
%FIGURE: DUST DISTRIBUTION R=30 vf=10
%%%%%%%%%%%%
\begin{figure*}
   \centering
   	\includegraphics[width=18.0cm]{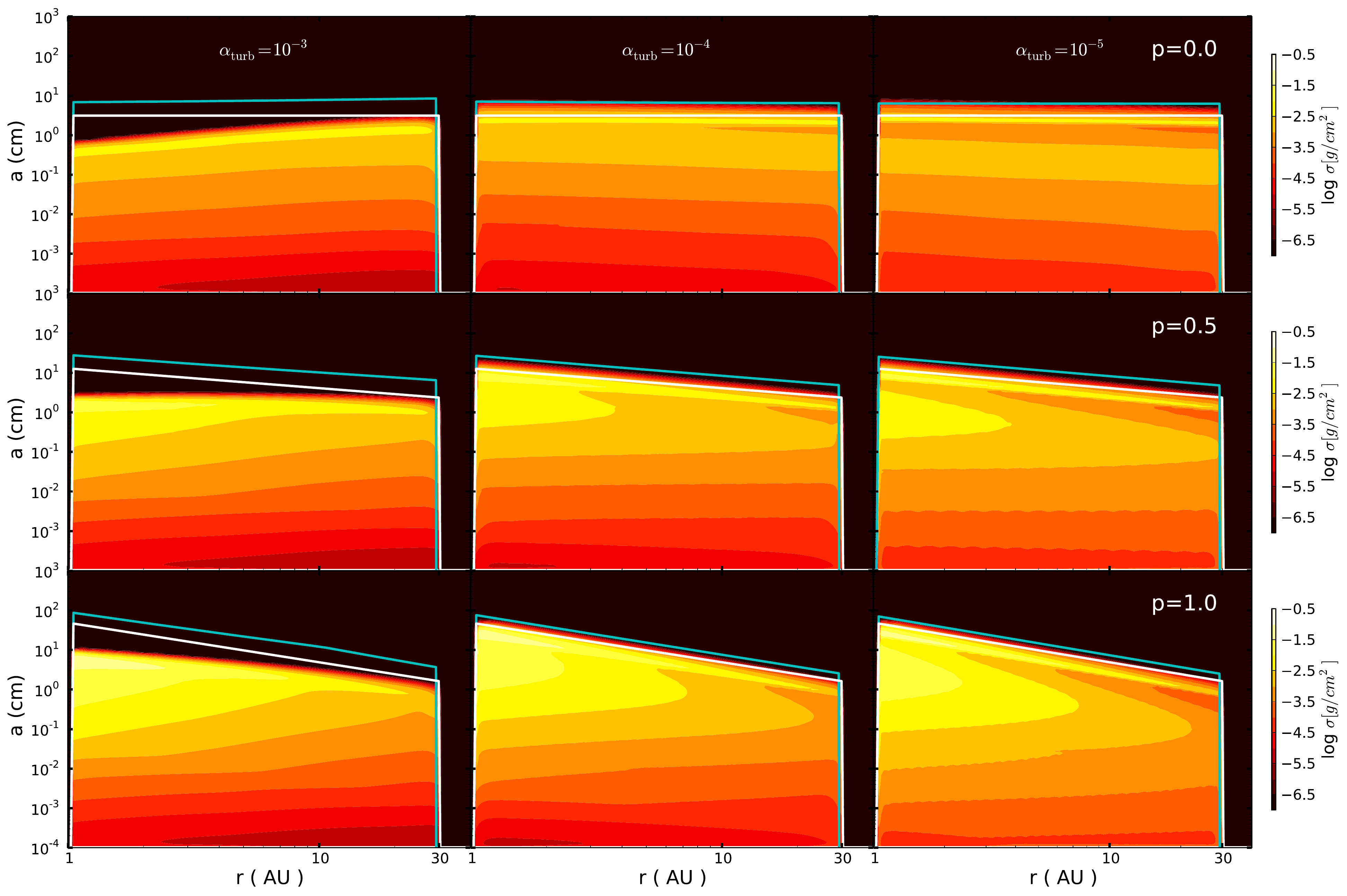}
   \caption{Vertically integrated dust density distribution after 1~Myr of evolution without  radial drift. Case of $R_{\mathrm{out}}~=~30$~AU and $v_f~=~10~$m~s$^{-1}$ and different values of the gas density slope $p~=~0$ \emph{(top panels)}, $p~=~0.5$ \emph{(middle panels)} and $p~=~1.0$  \emph{(bottom panels)}; with the turbulence parameter $\alpha_{\mathrm{turb}}~=~10^{-3}$  \emph{(left panels)}, $\alpha_{\mathrm{turb}}~=~10^{-4}$  \emph{(center panels)} and $\alpha_{\mathrm{turb}}~=~10^{-5}$  \emph{(right panels)}. The solid white line represents the particle size corresponding to St~=~1 (Eq.~\ref{eq:stokes}) and reflects the shape of the gas density, while the cyan line represents the maximum size that particles can reach given a fragmentation velocity $v_f$.}
   \label{dust_density_Rout30vf10}
\end{figure*}

\vskip 0.4cm
When we considered pressure inhomogeneities, we assumed the same prescription as in \cite{pinilla2012a}  to simulate long-lived pressure bumps, with the unperturbed density as a simple power law

\begin{equation}
	\Sigma_g' (r)= \Sigma_0~r^{-p}\left(1+A\cos\left[2\pi \frac{r}{L(r)}\right]\right),
  \label{eq:gas_bumpy}
\end{equation}

\noindent where the wavelength  of the perturbation $L(r)$ is taken to be a factor $f$ of the vertical disk scale-height $h(r)$ i.e. $L(r)~=~f~h(r)$, with 

\begin{equation}
	h(r)=c_s~\Omega^{-1}\propto \frac{L_{\star}^{1/8}}{\sqrt{M_{\star}}} \sqrt{r}.
	\label{eq:h}
\end{equation}

\cite{pinilla2012a} showed that under these assumptions, bumps with an amplitude of $\sim$~30\% and  a wavelength of one scale-height are necessary  for T-Tauri disks, to reduce the radial drift and keep the particles in the outer regions of the disk after several Myr. However, in Sect~\ref{sec2_2}, we showed that radial drift effects are stronger for BD disks, independently of the distance  of the particles from the star. In addition, since the wavelength of the perturbation was considered proportional to the disk scale-height $h(r)$,  at a given distance $r$, $L(r)$ would by definition be larger for BD than for T-Tauri disks (Eq.~\ref{eq:h}). This implies that for a given amplitude  of the perturbed density, the pressure gradient would be lower in BD disks, making the trapping of particles even more difficult. We used a fixed bump width equal to one scale height, i.e., $f~=~1.0$, since  the scale of turbulent structures from MRI are of the order of the scale height of the disk \citep{flock2011}, and they may be the origin of pressure inhomogeneities in the disk. The amplitudes considered are therefore higher than for T-Tauri disks and these are taken A=$\{0.4, 0.6\}$. All  stellar and disk parameters are summarized in Table~\ref{table_parameters}.

%%%%%%%%%%
\section{Results}     \label{sec:results} 
%%%%%%%%%%
In this section, we present the results of the numerical simulations setting the radial drift to zero,  followed by the case of non-zero radial drift and pressure bumps in the disk. For each case, the observational perspectives are also presented.

%%%%%%%%%%%%
%FIGURE: DUST DISTRIBUTION R=30 vf=30
%%%%%%%%%%%%
\begin{figure*}
   \centering
	\includegraphics[width=18.0cm]{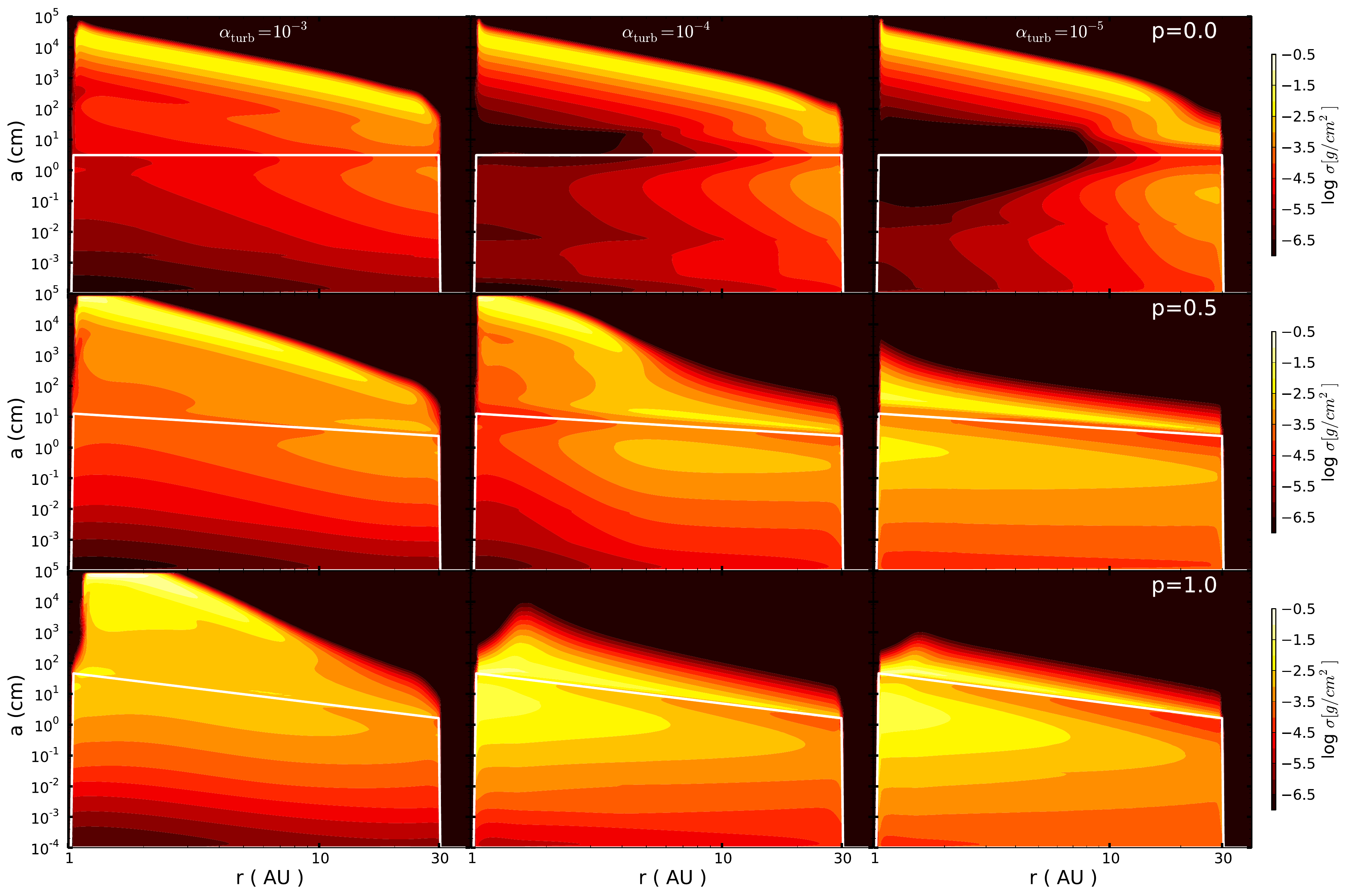}   
   \caption{Vertically integrated dust density distribution after 1~Myr of evolution without radial drift. Case of $R_{\mathrm{out}}~=~30$~AU and $v_f~=~30~$m~s$^{-1}$ and different values of the gas density slope $p~=~0$ \emph{(top panels)}, $p~=~0.5$ \emph{(middle panels)} and $p~=~1.0$  \emph{(bottom panels)}; with the turbulence parameter $\alpha_{\mathrm{turb}}~=~10^{-3}$  \emph{(left panels)}, $\alpha_{\mathrm{turb}}~=~10^{-4}$  \emph{(center panels)} and $\alpha_{\mathrm{turb}}~=~10^{-5}$  \emph{(right panels)}. The solid white line represents the particle size corresponding to St~=~1 (Eq.~\ref{eq:stokes}) and reflects the shape of the gas density.}
   \label{dust_density_Rout30vf30}
\end{figure*}

\subsection{No radial drift} \label{sec3_1} 

\subsubsection{Dust density distribution} \label{sec3_1_1}

Figures~\ref{dust_density_Rout30vf10} and \ref{dust_density_Rout30vf30} show the vertically integrated dust density distribution  (Eq.~\ref{eq:vertically}) after 1~Myr of dust evolution when the radial drift is set to zero, for the case of $R_\mathrm{out}~=~$30~AU and two different values of the fragmentation velocity: 10~$m~s^{-1}$ and 30~$m~s^{-1}$. In each case, the three different values of the gas density slope $p$ and the turbulence parameter $\alpha_{\mathrm{turb}}$ are plotted. For this case, the dust is considered to be in a steady state, therefore $u_{\mathrm{drag}}$ and $u_\mathrm{drift}$ are neglected in Eq.~\ref{eq:dustvel}.

\vskip 0.4cm
\emph{Effect of turbulence:} when particles grow and reach sizes such that they are more weakly coupled to the gas, the main sources of relative velocities are turbulent and azimuthal velocities, since radial drift velocities are set to zero. Because the highest turbulent relative velocities (Eq.~\ref{eq:u_turb}) depend on $\alpha_{\mathrm{turb}}$ and the disk temperature is assumed to be very low in BD disks ($\sim$~10K at 10~AU), fragmentation is not caused by turbulent motions. Instead, it mainly happens because the azimuthal dust velocities $u_{d, \phi}$ are as high as the limit beyond which particles fragment (fragmentation velocity, $v_f$). The azimuthal drift velocity $u_{d, \phi}$ is given by \citep{birnstiel2010a}

\begin{equation}
	u_{d, \phi}= \left | -\frac{\partial_r P}{\rho \Omega}\left(\frac{1}{1+\mathrm{St}^2}\right)\right |.
	\label{eq:u_azimuthal}
\end{equation}

Hence, as we noticed in Fig~\ref{dust_density_Rout30vf10}, the fragmentation limit  and consequently $a_{\mathrm{max}}$ are independent of the $\alpha_{\mathrm{turb}}$ and for a given gas density slope, the largest size that particles can reach is the same for each  $\alpha_{\mathrm{turb}}$ and only depends on the gas density profile. On the other hand, we notice that the concentration of mm- and cm-size grains is more evenly distributed in the whole disk for $\alpha_{\mathrm{turb}}~=~\{10^{-4}, 10^{-3}\}$ than for $\alpha_{\mathrm{turb}}~=~10^{-5}$, for which they are mainly in the inner part of the disk. 

The first remarkable difference of grain growth between T-Tauri and BD disks is that the turbulent mixing strength $\alpha_{\mathrm{turb}}$ does not play as important a role for dust fragmentation in BD disks as it does for T-Tauri disks. Without radial drift,  fragmentation in T-Tauri disks mainly occurs because of turbulent motion and the effect of different $\alpha_{\mathrm{turb}}$ values is significant \citep[see e.g ][for the effect of turbulence in the spectral index]{birnstiel2010b}, while for BD disks the azimuthal relative velocity dominates, which is independent of  $\alpha_{\mathrm{turb}}$.

%%%%%%%%%%%%
%FIGURE: Flux vs alpha.
%%%%%%%%%%%%
\begin{figure*}
   \centering
   \includegraphics[width=18.0cm]{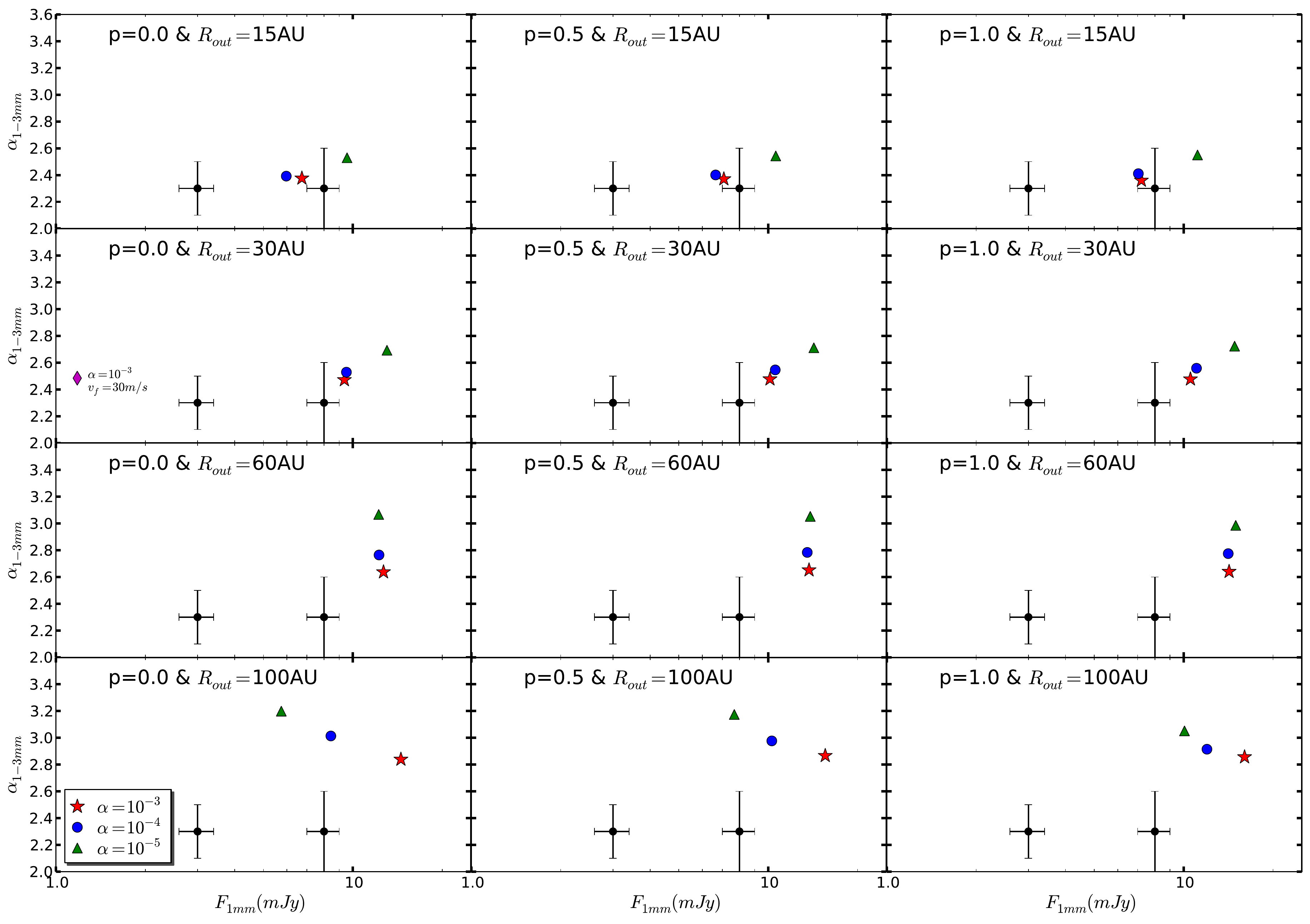}
   \caption{Predicted fluxes at ~1~mm ($F_{1\mathrm{mm}}$) and the spectral index between 1 and 3~mm ($\alpha_{\mathrm{1-3mm}}$) after 1~Myr of dust evolution, without radial drift, $v_f~=~10~m~s^{-1}$  and for all  other parameters discussed in Sect.~\ref{sec2_3}. Black dots with error bars are millimeter observations of the young BD $\rho$-Oph~102 \citep{ricci2012} and 2M0444+2512 \citep{ricci2013}}
   \label{flux_alpha}
\end{figure*}

\vskip 0.4cm

\emph{Effect of the gas density slope:}  For the same turbulence parameter and different  gas density slope $p$, the dust density distribution changes (see Figs.~\ref{dust_density_Rout30vf10} and \ref{dust_density_Rout30vf30}). When $p~=~1$  (steep surface density), large particles   ($a~>~$1~cm) are well coupled to the gas in the dense inner region and grow to even larger  sizes ($a~\sim~$50~cm) before they fragment. Therefore,  with $p~=~1$ the  mm-size particles are less concentrated  in the region where the gas density is lower, i.e., in the outer region of the disk ($r$~$\gtrsim$~10~AU), than for $p~=~\{0.5,1.0\}$. 

\vskip 0.4cm

\emph{Effect of the fragmentation velocity:} when the velocity at which particles fragment increases to 30~$m~s^{-1}$ (Fig.~\ref{dust_density_Rout30vf30}), grains can grow to very large sizes such that St~$>$~1. Fragmentation does not  occur and this implies that  bodies  that are no longer affected by gas ($a~\gtrsim~1$~m) grow and even in some cases reach the maximum size of particles  considered in these numerical simulations ($\sim$~1~km). Small particles are not replenished and as a result the disk can  be almost empty of small grains ($\lesssim$~1~cm). For example, when $p~=~0$  and $\alpha_{\mathrm{turb}}~=~10^{-5}$, most of the disk consists only of big grains ($a>10$~cm). However, when $p$ increases and the gas density is higher in the inner regions, larger particles are coupled to the gas and  can still be affected by gas turbulence. For this reason, there is some mm-dust in the inner part for the same $\alpha_{\mathrm{turb}}~=~10^{-5}$ and $p~=~1$.

As we discussed above, because of the low disk  temperatures, destructive collisions due to turbulence are less likely in BD disks, and even more with these high fragmentation velocities considered for ices (such as $v_f~=~30~m~s^{-1}$). With ices, the material properties vary, leading to fragmentation velocities of around $\sim~10-50~m~s^{-1}$ \citep{wada2009, wada2011}, but it is still a matter of debate if ices can have these high fragmentation velocities. Some scenarios produce disks with only large grains (Fig.~\ref{dust_density_Rout30vf30}), which may not  allow  a good agreement between these theoretical models and the recent submillimeter observations of disk around two BD. For this reason, we focus the following results on the case of $v_f~=~10~m~s^{-1}$, where the fragmentation allows  a constant replenishment of small particles.

\vskip 0.4cm

\emph{Effect of the outer radius:} When the outer radius of the disk decreases or increases, the same amount of dust is distributed in a smaller or larger region with the same $M_{\mathrm{disk}}$ and $\Sigma_g$ profile, and the possibility of having cm- mm-size particles in the outer regions changes, we discuss this effect below.

\subsubsection{Comparison with observations} \label{sec3_1_2} 

%%%%%%%%%%%%
%FIGURE: comparison between amplitudes 
%%%%%%%%%%%%
 \begin{figure*}
  \centering
   \begin{tabular}{cc}
   \includegraphics[width=8.5cm]{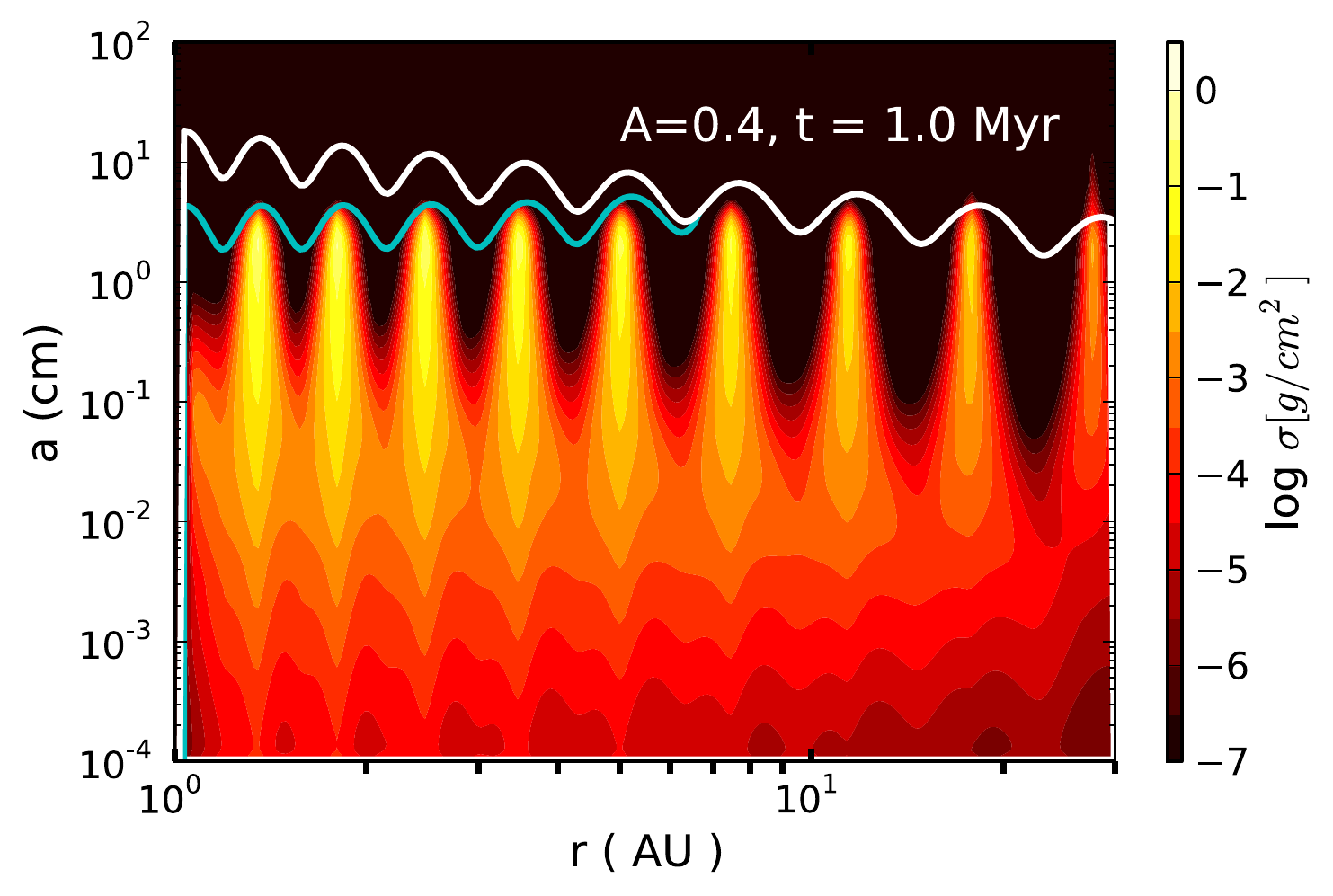} & 
      \includegraphics[width=8.5cm]{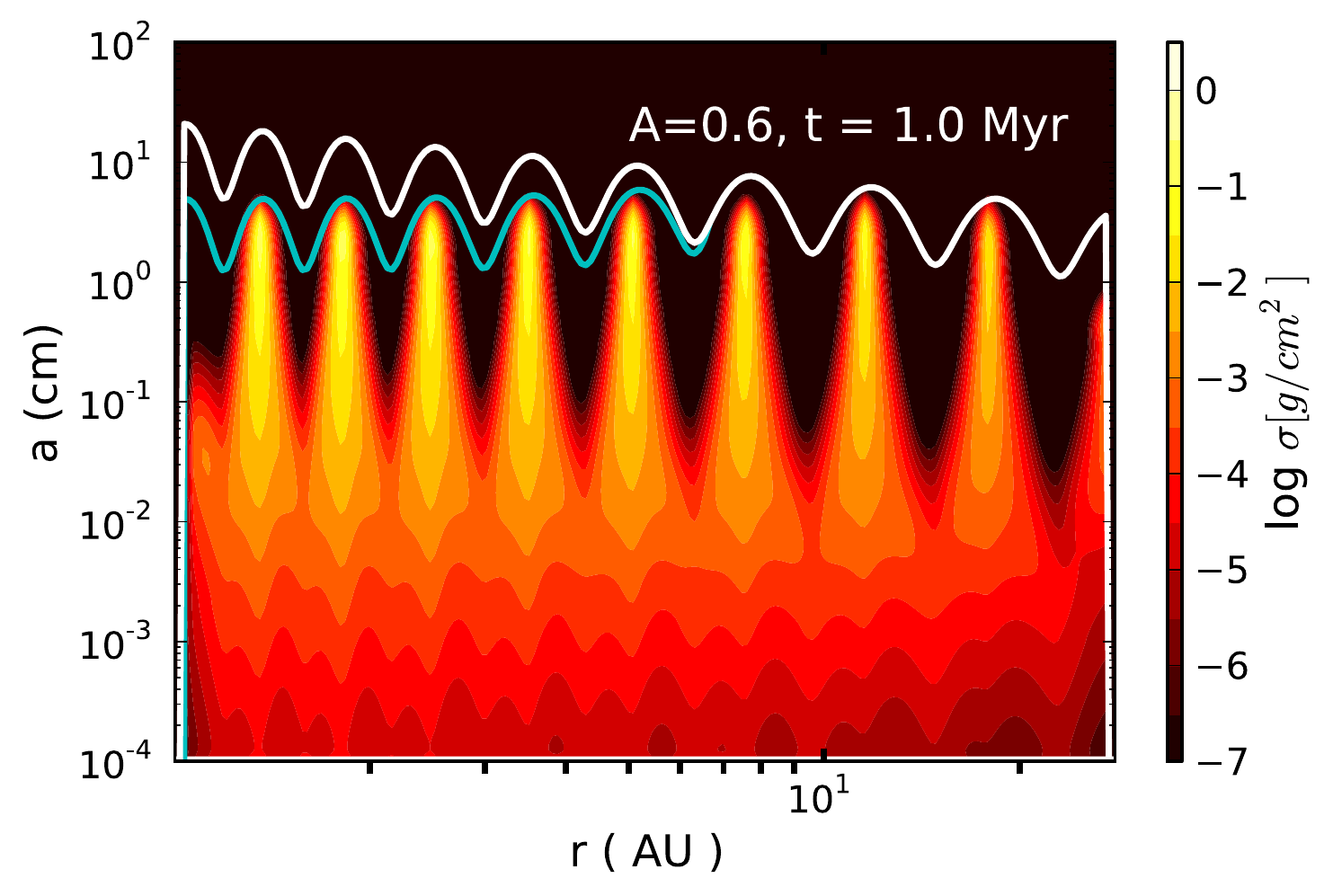}
    \end{tabular} 
   \caption{Vertically integrated dust density distribution after 5~Myr of evolution including radial drift and a bumpy gas surface density (Eq~\ref{eq:gas_bumpy}) with A~=~0.4 \emph{(left panel)} and A~=~0.6 \emph{(right panel)}. Case of $R_{\mathrm{out}}~=~30$~AU, $v_f~=~10~$m~s$^{-1}$, $p~=~0.5$ and $\alpha_{\mathrm{turb}}~=~10^{-3}$. The solid white line represents the particle size corresponding to St~=~1 (Eq.~\ref{eq:stokes}) and reflects the shape of the gas density, while the cyan line corresponds to the fragmentation limit.}
   \label{Fig:A04_and_A06}
\end{figure*}

To compare model predictions with disk fluxes of low-mass disks, we considered the dust distribution from the simulations after they almost reached a quasi-static state $\sim$~1~Myr.  We calculated the opacities $\kappa_\nu$ for each grain size and at a given frequency $\nu$. We used for simplicity optical constants for magnesium-iron silicates \citep{jaeger1994, dorschner1995} from the Jena database\footnotemark{}\footnotetext{$\textrm{http://www.astro.uni-jena.de/Laboratory/Database/databases.html}$} and followed the Mie theory. Once the opacities were calculated, the optical depth $\tau_\nu$ was computed as

\begin{equation}
	\tau_\nu=\frac{\sigma(r,a) \kappa_\nu}{\cos i},
  \label{eq:opt_depth}
\end{equation}

\noindent where $i$ is the disk inclination, which we took to be zero. The flux of the disk at a given frequency $F_\nu$ is therefore

\begin{equation}
	F_\nu=\frac{2\pi\cos{i}}{d^2}\int_{R_{\mathrm{in}}}^{R_{\mathrm{out}}} B_\nu(T(r)) [1-e^{-\tau_\nu}] r dr,
  \label{eq:flux}
\end{equation}

\noindent with $d$ being  the distance to the source, which was taken to be 140~pc as in the  young disks in Taurus and Ophiuchus star-forming regions. $B_\nu(T(r))$ is the Planck function for a given temperature profile $T(r)$.  At mm wavelengths, the flux is proportional to the dust mass in the outer region of the disks. Hence, the flux $F_\nu$ could be approximated as a power law $F_\nu\propto \nu^{\alpha_{\mathrm{mm}}}$, where the spectral index $\alpha_{\mathrm{mm}}$ gives information about the size of the grains and is expected to be lower than 3 when the dust reaches mm sizes \citep{natta2007}. Figure~\ref{flux_alpha} shows the predicted fluxes at ~1~mm ($F_{1\mathrm{mm}}$) and the spectral index between 1 and 3~mm ($\alpha_{\mathrm{1-3mm}}$) for $v_f~=~10~m~s^{-1}$ and all other parameters considered in Sect.~\ref{sec2_3}.

First of all, comparing $F_{1\mathrm{mm}}$ and $\alpha_{\mathrm{1-3mm}}$ for different $R_{\mathrm{out}}$, we notice that the spectral slope increases when the disk is more extended. This is a natural result of distributing initially the same amount of dust in an extended region, decreasing the possibility of having mm-size grains. The obtained fluxes are between $\sim$~1-20 mJy for all cases. In addition,  a single case (the only diamond-point of Fig.~\ref{flux_alpha}) where fragmentation does not happen is plotted, with $v_f~=~30~m~s^{-1}$, $R_{\mathrm{out}}$~=~30~AU and $p~=~0.0$. It is possible to see that most of the dust in the disk has grown to large sizes (a~$>$~10~cm). When the particles have such a large size, they have very low opacities, which results in low millimeter fluxes. The predicted fluxes are very low (close to ~1~mJy) compared with the other fluxes (see top-left panel of Fig.~\ref{dust_density_Rout30vf30} and the corresponding diamond-point in the second-left panel from the top to the bottom of Fig.~\ref{flux_alpha}). 

In general, when fragmentation still occurs, i.e., for $v_f~=~10m~s^{-1}$, the integrated flux weakly depends on the gas density slope $p$, therefore we focus in the subsequent results on the intermediate value, i. e., $p~=~0.5$.  In most of the cases, the spectral slope and the flux are slightly sensitive to the turbulence parameter $\alpha_{\mathrm{turb}}$, and $\alpha_{\mathrm{1-3mm}}$ increases for low turbulence. This is because there are fewer mm-size grains that contribute to decrease the spectral index when $\alpha_{\mathrm{turb}}$ is low (Fig.~\ref{dust_density_Rout30vf10} and Fig.~\ref{dust_density_Rout30vf30}), as was discussed in Sect.~\ref{sec3_1_1}. 

Comparing the millimeter observations (dot points with error bars in Fig~\ref{flux_alpha}) of two BD disks (2M0444+2512 and $\rho$ Oph 102), it is important to notice that when the radial drift is set to zero,  the predicted and the observed values of $F_{1\mathrm{mm}}$ and $\alpha_{\mathrm{1-3mm}}$ match well for the brightest BD disk and in particular for $R_{\mathrm{out}}~=~15~$AU.  The error bars come from the the optical depth uncertainties for different grain composition \citep{beckwith1990}. When the extension of the disk increases to $R_{\mathrm{out}}~=~30~$AU, the spectral slope has values that agree with observations (especially for higher values of $\alpha_{\mathrm{turb}}$), but the fluxes are slightly higher than those detected by mm-observations. This leads to the conclusion that in this case, a combination of fragmentation with a minor drift is necessary to reduce the number of mm-grains and have a better agreement between theoretical predictions and recent millimeter observations of 2M0444+2512. For the faintest disk $\rho$ Oph 102, radial drift is indeed needed in any case. For $R_{\mathrm{out}}~=~60~$AU, only few cases allow to have $\alpha_{\mathrm{1-3mm}}~\lesssim~3$, nevertheless, these cases  match the millimeter observations less well . Finally, for most of the cases of $R_{\mathrm{out}}~=~100~$AU,  the model predictions for the millimeter fluxes and spectral indices  are inadequate to explain the observations.

As a conclusion, theoretical models of dust evolution in which the radial drift is set to zero and considering  BD disk conditions such as low-mass disks ($M_{\mathrm{disk}}~=~2~M_{\mathrm{Jup}}$), low radial extension ($R_{\mathrm{out}}~=~15~$AU), presence of ices that allow fragmentation velocities  of about $v_f~=~10~m~s^{-1}$ and average turbulence strength $\alpha_{\mathrm{turb}}~=~\{10^{-4}, 10^{-3}\}$, are the models with the best agreement to mm-observations.

\subsection{Radial drift and pressure bumps} \label{sec3_2}

We focus our attention on the  most favorable cases: $R_{\mathrm{out}}~=~\{15, 30, 60\}$~AU (particularly  15 and 30~AU) and $\alpha_{\mathrm{turb}}~=~10^{-3}$. The gas density slope was taken to be $p~=~0.5$. The amplitudes for the long-lived pressure bumps were considered $A~=~\{0.4, 0.6\}$ (Eq.~\ref{eq:gas_bumpy}). \cite{pinilla2012a} showed that these  inhomogeneities are comparable with global simulations of zonal flows with an amplitude of $\sim$~25\% by \cite{uribe2011}. 

For the simulations with radial drift, $u_{\mathrm{drag}}$ and $u_{\mathrm{drift}}$ for Eq.~\ref{eq:dustvel} were taken into account. This implies that radial drift also contributes to the total relative velocities of dust particles.

\subsubsection{Dust density distribution}

In a region where the pressure gradient is positive, gas moves with super-Keplerian velocity and particles with sizes corresponding to St~$\sim~1$  would move outward. As a consequence, inside the pressure bumps  the amount of dust naturally increases, allowing an increase in the frequency of sticking collisions. Figure~\ref{Fig:A04_and_A06} compares the dust density distribution after 1~Myr of evolution for the case of $R_{\mathrm{out}}~=~30$~AU, $v_f~=~10~$m~s$^{-1}$, $p~=~0.5$, $\alpha_{\mathrm{turb}}~=~10^{-3}$ and two different values of the amplitude of the perturbation $A~=~0.4$ and $A~=~0.6$. In both cases, with the pressure bumps considered, it is possible to reduce the strong radial drift and retain mm-size particles even after million-year timescales.  The efficiency of   the pressure bumps is evident, with many more mm grains than in the test case of Fig.~~\ref{with_drift} (empty disk after 0.5~Myr of dust evolution, Fig.~\ref{with_drift}). In addition, the effectiveness of the amplitude strength, i.e., the pressure gradient strength, for the trapping of particles is almost the same for both amplitudes after 1~My of evolution.

%%%%%%%%%%%%
%FIGURE: dust-to-gas-ratio 
%%%%%%%%%%%%
\begin{figure}
   \centering
	\includegraphics[width=8.0cm]{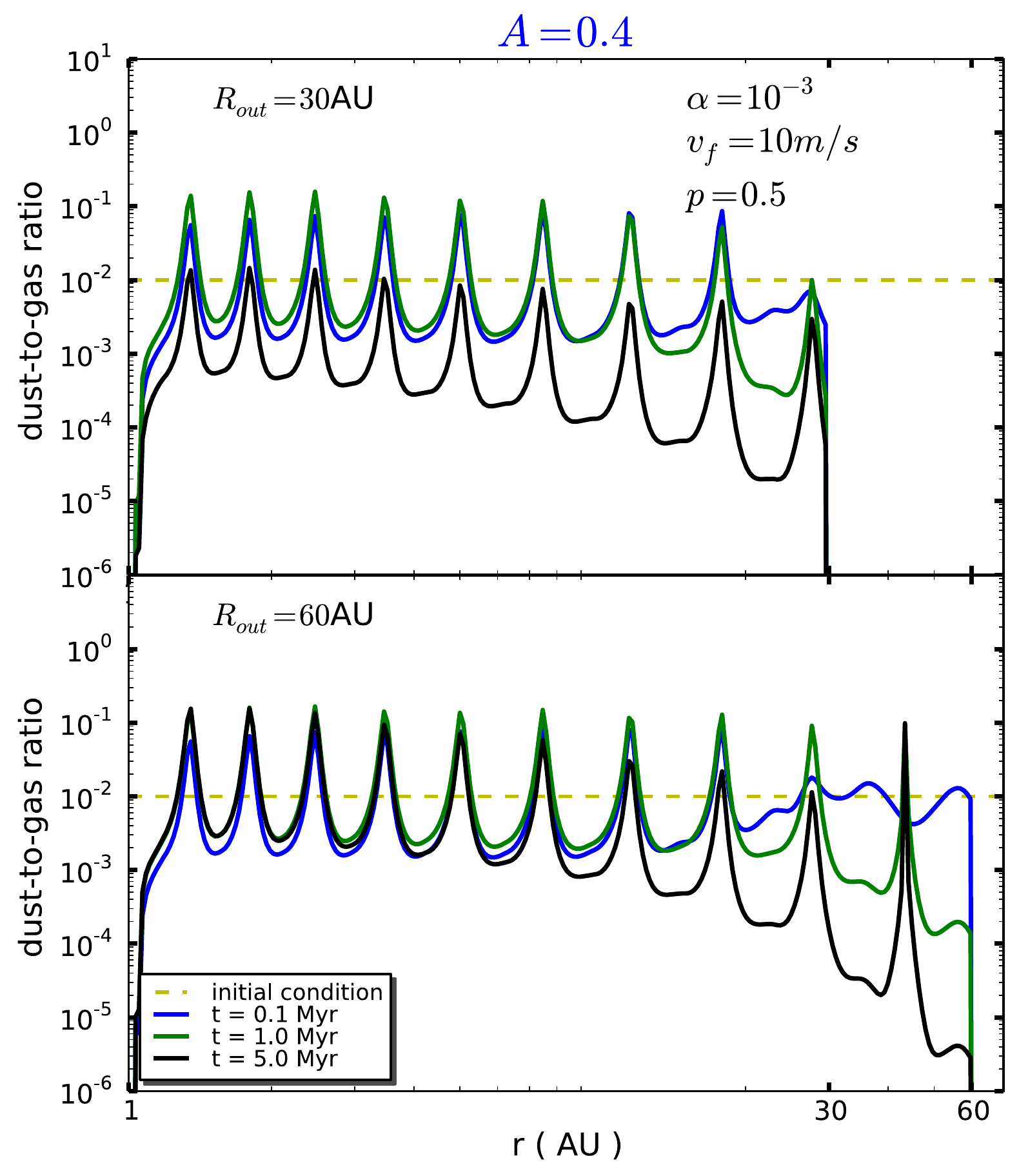}
   \caption{Dust-to-gas ratio at different times of dust evolution including radial drift and a bumpy gas surface density (Eq~\ref{eq:gas_bumpy}) with  A~=~0.4. Case of  $\alpha_{\mathrm{turb}}~=~10^{-3}$, $v_f~=~10~$m~s$^{-1}$, $p~=~0.5$,     $R_{\mathrm{out}}~=~30$~AU \emph{(top panel)} and $R_{\mathrm{out}}~=~60$~AU \emph{(bottom panel)}.}
   \label{comparison_ratio}
\end{figure}

Inside the bumps, the radial drift is reduced and fragmentation is also less efficient than in the case of  Fig.~\ref{with_drift}. With a fragmentation velocity of $v_f~=~10~m~s^{-1}$, particles grow to sizes corresponding to St~$\sim~1$ for the cases in Fig.~\ref{Fig:A04_and_A06}, and because of the positive pressure gradient, grains are trapped inside the bumps. However, since fragmentation  due to turbulent and azimuthal relative velocities is still happening, when particles collide and become smaller such that St~$<$~1, they are more difficult to trap in the pressure bump \citep{pinilla2012a} and  would finally drift to the star. 

%%%%%%%%%%%%
%FIGURE: time evolution of flux_alpha
%%%%%%%%%%%%
\begin{figure*}
   \centering
	\includegraphics[width=18.0cm]{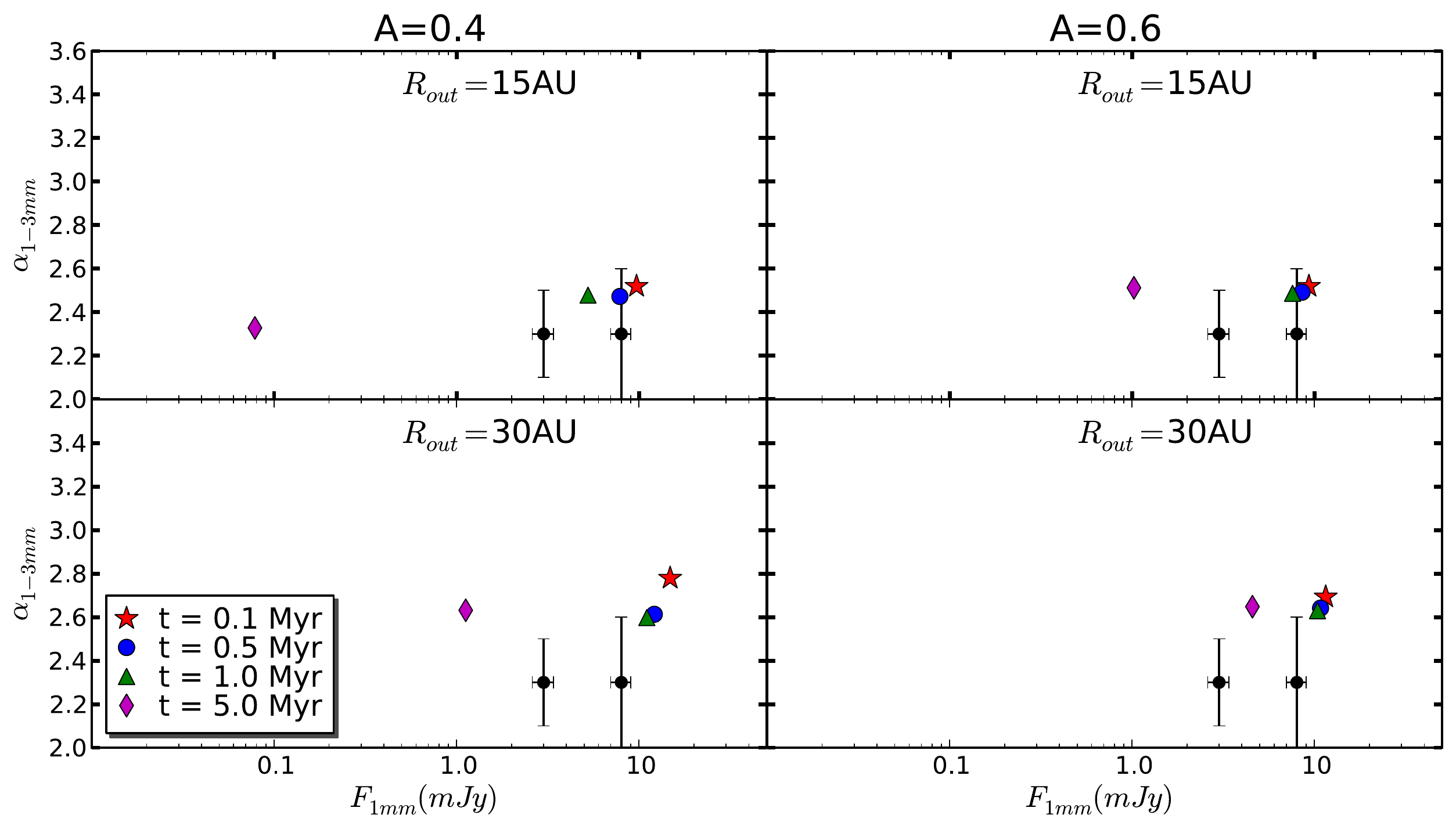}
   \caption{Time evolution of predicted fluxes at ~1~mm ($F_{1\mathrm{mm}}$) and the spectral index between 1 and 3~mm ($\alpha_{\mathrm{1-3mm}}$) including radial drift and a bumpy gas surface density (Eq.~\ref{eq:gas_bumpy}) with $R_{\mathrm{out}}~=~\{15, 30\}~$AU, $\alpha_{\mathrm{turb}}~=~10^{-3}$, $v_f~=~10~m~s^{-1}$, A~=~0.4 \emph{(left panels)}, A~=~0.6 \emph{(right panels)}, and $p~=~0.5$. Black dots with error bars are millimeter observations of the young BD $\rho$-Oph~102 \citep{ricci2012} and 2M0444+2512 \citep{ricci2013}.}
   \label{evolution_flux_alpha}
\end{figure*}

\subsubsection{Dust-to-gas ratio}

Figure~\ref{comparison_ratio} shows the radial dependence of the dust-to-gas ratio at different times of evolution for the cases of $\alpha_{\mathrm{turb}}~=~10^{-3}$, $v_f~=~10~$m~s$^{-1}$, and  $p~=~0.5$, in a bumpy gas surface density with $A~=~0.4$ and $R_{\mathrm{out}}~=~\{30, 60\}$~AU. For this case, the dust-to-gas ratio remains almost constant with time until 1~Myr of evolution. It increases inside the pressure bumps and varies between $\sim~10^{-2}-10^{-1}$, but there is a remarkable reduction after 5~Myr of evolution, where it can reach values of $10^{-3}$ in the inner part and $10^{-4}-10^{-5}$ in the outer parts of the disk. 

When the outer radius increases, the dust-to-gas ratio does not remain constant with time in the outer regions of the disk, and it is possible to distinguish that it decreases with time, reaching even values of $\sim~10^{-6}$ in those regions (bottom panel of Fig.~\ref{comparison_ratio}).

It is important to note that the gas density profile remains constant for our dust evolution models, since viscous evolution timescales are longer than dust growth timescales. In addition,  we did not consider any mechanism that might selectively disperse the gas, such as photoevaporation \citep[see e.g.][]{gorti2009, owen11}.

\subsubsection{Comparison with millimeter observations} \label{sec3_2_1} 

Focusing the attention on the cases where fragmentation is effective (i.e., when $v_f~=~10~m~s^{-1}$ and $\alpha_{\mathrm{turb}}~=~10^{-3}$), Fig.~\ref{evolution_flux_alpha} shows the time evolution of $F_{1\mathrm{mm}}$ and $\alpha_{\mathrm{1-3mm}}$ including radial drift and a bumpy gas surface density  with $R_{\mathrm{out}}~=~\{15, 30\}~$AU,  $p~=~0.5$, and A~=~$\{0.4,0.6\}$. For $A~=~0.4$, we notice a good match between the theoretical predictions and the observations, even after 1~Myr of evolution, there are still enough mm-grains in the outer regions that contribute to have high values of the flux at 1~mm. The fluxes become lower at 5~Myr, because a significant amount of grains continues to fragment and drift toward the star. However,  $F_{1\mathrm{mm}}$ and $\alpha_{\mathrm{1-3mm}}$ at intermediate time steps like 1~-~2~Myr coincide well with the observed values, and  are in the range of the estimated ages of the observed BD disks. The effect of increasing the pressure gradient, i.e.,   higher $A$, is notable, since the fluxes increase because there are more trapped grains, in particular at long timescales, for instance 5~Myr.

\begin{figure}
   \centering
	\includegraphics[width=8.5cm]{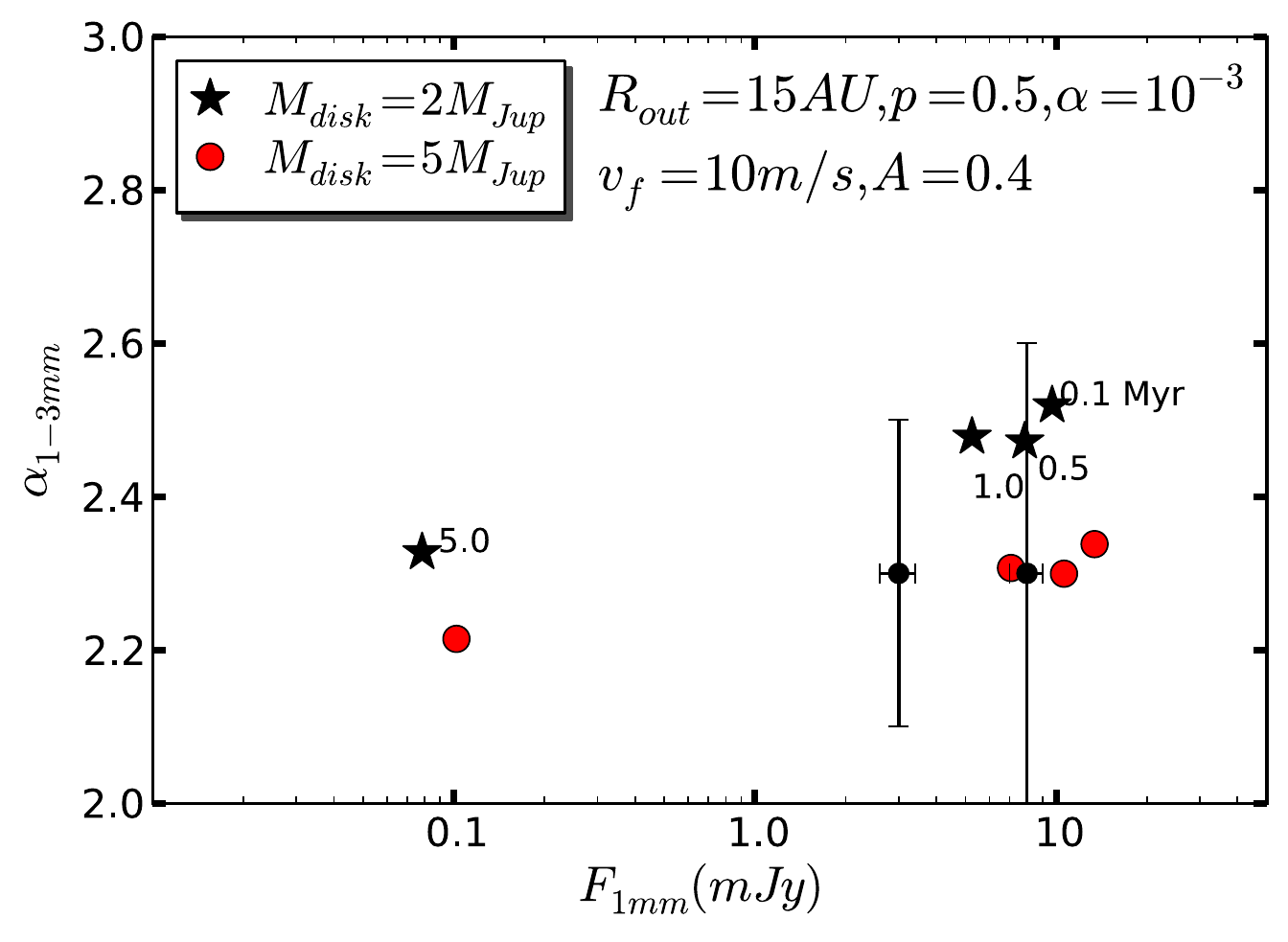}
   \caption{Comparison  of the predicted fluxes at ~1~mm ($F_{1\mathrm{mm}}$) and the spectral index between 1 and 3~mm ($\alpha_{\mathrm{1-3mm}}$) for two different disk masses ($M_{\mathrm{disk}}~=~2~M_{\mathrm{Jup}}$-star points and $M_{\mathrm{disk}}~=~5~M_{\mathrm{Jup}}$-dot points).  Time-evolving data from right to  left (from 0.1-5.0~Myr as Fig.~\ref{evolution_flux_alpha}). Case of $R_{\mathrm{out}}~=~15$~AU, $p=0.5$, $\alpha_{\mathrm{turb}}~=~10^{-3}$, $v_f~=~10~m~s^{-1}$, and $A~=~0.4$.}
   \label{comparison_massive_BD}
\end{figure}

To achieve a better agreement of the predicted fluxes $F_{1\mathrm{mm}}$ and the spectral index $\alpha_{\mathrm{1-3mm}}$ with the current mm-observations of 2M0444+2512 and $\rho$ Oph 102, it would be necessary to increase the mass of the disk. Figure~\ref{comparison_massive_BD} shows the comparison  of $F_{1\mathrm{mm}}$ and the $\alpha_{\mathrm{1-3mm}}$ for two different disk masses ($M_{\mathrm{disk}}~=~\{2, 5\}~M_{\mathrm{Jup}}$) and the specific case of $R_{\mathrm{out}}~=~15$~AU, $p=0.5$, $\alpha_{\mathrm{turb}}~=~10^{-3}$, $v_f~=~10~m~s^{-1}$ and $A~=~0.4$. Because there is more dust for $M_{\mathrm{disk}}~=~5~M_{\mathrm{Jup}}$ than for $M_{\mathrm{disk}}~=~2~M_{\mathrm{Jup}}$, there are more mm-grains that contribute to the integrated flux, and therefore the spectral slope can be lower.

%%%%%%%%%%
\section{Summary and discussion}     \label{sec:discussion}
%%%%%%%%%%
One of the most important problems in the core accretion theory for planet formation is ``the meter-size barrier''. This phenomenon results from the combination of high radial drift and fragmentation. A meter-sized object at 1~AU drifts toward the star on timescales much shorter than the growth timescales. In the colder regions of a disk, the same problem occurs when translated to mm-sized grains.  Both radial drift and fragmentation   appear to be different for particles in BD disks. 
  
We have studied how dust evolves and explored different scenarios where mm-grains survive in BD disks. For the results presented here, we considered typical BD parameters. For instance, the disk outer regions reach temperatures of $\sim$~10~K. Assuming that MRI is the source of turbulence, which  depends basically on the disk temperature and considering $\alpha_{\mathrm{turb}}$-prescription, low  and moderate values of $\alpha_{\mathrm{turb}}$ were taken ($10^{-5}~-~10^{-3}$). In addition, the presence of ices in the disk was assumed and fragmentation velocities were taken to be  $v_f~=~\{10,30\}~m~s^{-1}$, based on  numerical simulations of collisions with ices \citep[see e.g.][]{wada2009}. Furthermore, we considered low-mass disks ($2~M_{\mathrm{Jup}}$) and typical values for masses and luminosities of BD (see Table~\ref{table_parameters}). For the radial extension of BD disks, we considered $R_{\mathrm{out}}~=~\{15, 30,60,100\}$~AU. \cite{luhman2007} presented observations combining Spitzer spectroscopy and high-resolution imaging from HST of a circumstellar disk that is inclined close to edge-on around a young brown dwarf in Taurus, estimating a disk radius of $\sim~$20-40~AU. ALMA-Cycle~0 observations of $\rho$-Oph~102  \citep{ricci2012} do not resolve the disk, but they estimate an outer radius for the dust of $\sim$15-40~AU. From 3-mm CARMA observations of 2M0444+2512 \citep{ricci2013}, the outer radius for the dust is estimated to be $\sim$~20-75~AU for $p~=~1$ and $\sim$~15-30~AU for $p~=~0.5$.

%%%%%%%%%%%%
%FIGURE: Comparison mass of the disk
%%%%%%%%%%%%

\subsection{Fragmentation and drift barriers}

We have shown how several dust growth aspects in BD disks differ from T-Tauri and  Herbig Ae/Be disks. In the first stages of dust evolution, when particles  move along with the gas and coagulate, relative velocities are mainly due to Brownian motion and settling to the midplane. It was discussed in Sect.~\ref{sec2_2} that growth timescales due to settling are shorter at the location of a given temperature in BD than in T-Tauri disks.

When particles reach sizes such that they start to decouple from the gas (i.e. St~$\sim~$1), radial drift and turbulent motion become the main sources for relative velocities. On one hand, we demonstrated in Sect.~\ref{sec2_2} that radial drift can be even twice as fast for particles in BD disks than in T-Tauri disks, and as a consequence they would be depleted on shorter timescales. On the other hand, considering moderate  $\alpha_{\mathrm{turb}}$ and low disk temperatures, turbulent relative velocities are quite low. If turbulence is the main cause for fragmentation, destructive collisions are less likely in BD disks than in T-Tauri disks. When radial drift and fragmentation are considered, BD disks would be dust-poor after only few Myr because of the fast inward migration of dust grains (Fig.~\ref{with_drift}).

\subsection{No radial drift}
This drastic case, where radial drift is not included, is important  for studying whether mm-grain can be indeed formed considering dust coagulation/fragmentation models. When radial drift was set to zero, we found that two different scenarios can happen under BD disk conditions. In one scenario fragmentation occurs because of the relative azimuthal velocities, considering $v_f~=~10~m~s^{-1}$ (Fig.~\ref{dust_density_Rout30vf10}). In this case, it is possible to form mm-size particles in BD disk and achieve a good match for the predicted mm-fluxes and spectral indices, in particular for the brightest BD disk  2M0444+2512 (Fig.~\ref{flux_alpha}) for $R_{\mathrm{out}}~=~15~$AU. However, for larger disk sizes, i.e., $R_{\mathrm{out}}~=~\{30, 60\}~$AU, a slight radial drift is necessary to decrease the predicted mm-fluxes. For the other BD disk $\rho$ Oph 102, radial drift is always needed to reduce  the mm-fluxes by at least a factor of 3.

The second scenario is  when fragmentation is less likely, i.e., $v_f~=~30~m~s^{-1}$ (see e.g. Fig.~\ref{dust_density_Rout30vf30}), when most of the disk consists  of big grains (a~$>$~10~cm), leading to low mm-fluxes that underestimate millimeter observations. Therefore, values such as $v_f~=~30~m~s^{-1}$ do not yield a good agreement between models and observations, and average values of $v_f~=~10~m~s^{-1}$ are needed.

\subsection{Radial drift and pressure bumps}
To reduce the rapid inward drift that dust particles experience in protoplanetary disks and explain the presence of mm-grains in the outer regions of disks, pressure bumps have been suggested as a possible solution \citep[see e.g.][]{pinilla2012a}. In this work, when radial drift was taken into account,  we considered strong gas density inhomogeneities (40\% and 60\% of amplitude), that led to regions in the disk with a positive pressure gradient. These  pressure bumps could be the result of MRI effects \citep{johansen2009, uribe2011}. However, there are no known mechanisms that could produce this type of pressure bumps with these high amplitudes  globally present in the disk. Nevertheless, if the scale-height of a disk is higher, as in the case of BD disks, the scale of turbulent structures increases, leading to higher pressure scales \citep[e.g.][]{flock2011}.

In addition to global pressure inhomogeneities from MRI in the disk, a single and high-pressure bump can exist in disks. For instance, Rossby  wave instability (RWI) generated by the presence of a dead zone \citep{regaly2012} would create a high single-pressure bump. The possibility of a dead zone in a BD disk is still  debated. On the other hand,  the presence of a massive planet could be also the reason for trapping of particles in a single huge pressure bump \citep{pinilla2012b}. However, from low-mass disks, it is unlikely to have planets massive enough to open a gap. The criterion to open a gap in a disk depends on the local scale height $h(r)$, meaning that a clear gap would be less likely for BD than T-tauri disks for the same disk and planet parameters \citep{crida2006}. Hence, these two possibilities were ruled out for the conditions that we  considered for BD disks, and we only focused  on global-synthetic pressure bumps that allow  a broad radial distribution of millimeter dust grains in the whole disk. 

For simulations with pressure bumps, we considered a sinusoidal perturbation for the gas surface density with two different amplitudes A~=~$\{0.4,0.6\}$ (Eq~\ref{eq:gas_bumpy}) and a wavelength equal to one disk scale-height. The parameters considered for these simulations were taken from the most optimistic no-drift cases. The pressure bumps cause an accumulation of dust in the location of pressure maxima, where radial drift is reduced and fragmentation due to radial drift decreases. Inside those regions, fragmentation would happen only because turbulent or azimuthal motions and a good match between theoretical perspectives and spectral slopes found with millimeter observations for the two BD disks is possible, especially for $R_{\mathrm{out}}~=~\{15, 30\}~$AU. Increasing the mass of the disk from 2 to 5~$M_{\mathrm{Jup}}$, theoretical predictions and mm-observations match even better. The disk mass predictions for $\rho$-Oph~102 \citep{ricci2012} are 0.2-0 .6~$M_{\mathrm{Jup}}$ and for 2M0444+2512 \citep{ricci2013} are 2.0-5.0~$M_{\mathrm{Jup}}$, assuming a gas-to-dust mass ratio of 100.\\

\subsection{Comparison between T-Tauri and BD disks}
Although T-Tauri and BD disks have very different properties, a first-order comparison between the models which agree well with millimeter observations in each case can be made.  
If we compare the best model described in \cite{pinilla2012a} for T-Tauri disks (pressure bumps with an amplitude $A$ of 30\% and one scale-height of the disk as the width $f$ of the bumps) with the best model for BD disks ($A=0.4$, $f=1.0$, $R_{\mathrm{out}}=15$~AU and $M_{\mathrm{disk}}=2~M_{\mathrm{Jup}}$), considering the same disk and dust properties (fragmentation velocity ($v_f=10$m~s$^{-1}$), viscosity ($\alpha_{\mathrm{turb}}=10^{-3}$), initial dust-to-gas ratio(1/100), etc), we find that 85\% of the dust mass is in grains larger than 1~mm in the BD case, versus  70\% for T-Tauri disks. As a consequence, we expect our models to produce slightly flatter SEDs for BD disks than for TTS disks. We stress that this conclusion, which is inferred from our 1-D models that are only relevant in the mid-plane, should be taken with care. A proper vertical structure calculation is needed to assess this question, and see if the effect is strong enough to account for 15-20\% variation reported by e.g. \cite{mulders2012} on Chameleon~I targets.

\subsection{Additional improvements}  \label{sec4_2}
%%%

The aim of this paper was to determine the best conditions for grain growth in the mid-plane of BD disks. Our 1-D model is well-suited to this goal, but requires to azimuthally average the surface density profiles, which implies that we do not take into account possible local disk features that may locally affect dust growth.  2-D and 3-D calculations would allow the vertical disk structure and vertical grain distribution to be computed and  provide us predictions at shorter wavelengths (e.g. shape of the SED and of the silicate feature) that can be directly compared with large surveys.

Our model can also be improved in the prescription of the dust-evolution process. Recent laboratory experiments using silicates have revealed that particles probably also bounce  \citep{Guttler2010}, which is not included in our model so far. It is  an open question however whether this happens in the case of collision between ices \citep{wada2011}, and whether it has a strong impact on dust growth. In fact, the bouncing barrier \citep{Zsom2010} can be overcome if Maxwellian velocity distributions of relative velocities among dust grains are considered \citep{windmark20012a}. \cite{windmark2012} also showed that after including bouncing effects and considering mass-transfer collisions, km-sized objects can still form at few AU from the star, but only if radial drift is neglected and  cm-sized seeds are inserted initially.

Finally, the main caveat of our study is that we assumed that the gas density profile remains static for million-year timescales, neglecting the impact of different physical processes,  for example  the potential  time evolution of the  pressure bumps, like the case of zonal flows \citep[see e.g.\,][]{uribe2011}. Simultaneous modeling of gas and dust evolution is a step forward of our models.

\section{Conclusion}     \label{sec:conclusion_BD}
The first steps of planet formation differ between disks around T-Tauri stars and disks around BD. In BD disks, settling is more efficient and they are expected to be flatter. In addition, when dust grows and particles start to decouple from the gas, radial drift is a major problem and fragmentation due to turbulence is less likely than in T-Tauri disks. Our models show that, when BD disks are small (e.g., $R_{\mathrm{out}}\lesssim$~15AU), radial drift needs to be completely suppressed to account the mm-observations of one of the two BD disks observed so far. When disks are larger,  a small degree of radial drift is necessary for both cases. Including radial drift,  the most favorable conditions after million-year timescales of dust evolution are with  strong pressure inhomogeneities of an amplitude of $\sim$~40\%-60\%. Global dust evolution models that include the vertical structure are the next step to reproduce all observational signatures of BD disks,  for example the flatter geometry of BD compared with T-Tauri disks, or the large average grain sizes in BD disk atmospheres.

\begin{acknowledgements}

We thank the anonymous referee for his/her constructive report, which helped us to improve and clarify the main results of the paper. T.~Birnstiel acknowledges support from NASA Origins of Solar Systems grant NNX12AJ04G. P.~Pinilla acknowledges the CPU time for running simulations in bwGRiD, member of the German D-Grid initiative, funded by the Ministry for Education and Research (Bundesministerium f\"ur Bildung und Forschung) and the Ministry for Science, Research and Arts Baden-Wuerttemberg (Ministerium f\"ur Wissenschaft, Forschung und Kunst Baden-W\"urttemberg).

\end{acknowledgements}

%%%%%%%%%%%%%
%% REFERENCES
%%%%%%%%%%%%%

\bibliographystyle{aa}  
\bibliography{Pinilla_brown_dwarfs.bbl}

%%%% END%%%%%%%

\end{document}